\documentclass{aa}

\usepackage{graphicx}

\usepackage{graphicx}

\newcommand{\swift}{{\it Swift}}
\newcommand{\xmm}{{\it XMM-Newton}}
\newcommand{\rosat}{{\it ROSAT}}
\newcommand{\kepler}{{\it Kepler}}

\newcommand{\rxte}{{\it RXTE}}
\newcommand{\nustar}{{\it NuSTAR}}

\begin{document}

\title{\xmm\ observation of V1504\,Cyg as a probe for the existence of an evaporated corona}
\titlerunning{\xmm\ observation of V1504\,Cyg}
\authorrunning{Dobrotka et al.}

\author{A.~Dobrotka \inst {1}, J.-U.~Ness \inst {2} A.A.~Nucita \inst {3,4} and M.~Melicher\v{c}\'ik \inst {1}}

\offprints{A.~Dobrotka, \email{andrej.dobrotka@stuba.sk}}

\institute{Advanced Technologies Research Institute, Faculty of Materials Science and Technology in Trnava, Slovak University of Technology in Bratislava, Bottova 25, 917 24 Trnava, Slovakia
\and
			XMM-Newton Science Operations Centre, European Space Astronomy Centre, Camino Bajo del Castillo s/n, Urb. Villafranca del Castillo, 28692 Villanueva de la Ca\~nada, Madrid, Spain
\and
			Department of Mathematics and Physics E. De Giorgi, University of Salento, via per Arnesano, CP 193, 73100, Lecce, Italy
\and
			INFN, Sez. di Lecce, via per Arnesano, CP 193, 73100, Lecce, Italy
}

\date{Received / Accepted}

\abstract
% context heading (optional)
% {} leave it empty if necessary
{}
% aims heading (mandatory)
{We present an analysis of an \xmm\ observation of the dwarf novae V1504\,Cyg during the decline from an outburst. Our goal is to search for evidence for an evaporated X-ray corona. Such a corona can be understood as an optically thin geometrically thick disc around a central part of an optically thick geometrically thin disc.}
% methods heading (mandatory)
{We study the X-ray spectra using a cooling flow model and the evolution of the amplitude of variability and power density spectra in UV and X-rays.}
% results heading (mandatory)
{The X-ray (pn) count rate increases from initially around 0.03 cps to 0.17 cps with a harder spectrum and a higher degree of variability. Meanwhile, the OM/UVW1 light curve follows a slow decline with decreasing amplitude of variability. For further study we split the X-ray data into two parts, and analysed them separately. Both parts are described by a cooling flow model, while the first low luminosity part requires an additional power law component suggesting presence of a wind. Spectral fitting revealed a higher temperature during the second brighter part. Timing analysis reveals a potential break frequency at log($f$/Hz) = -3.02 during decline towards the quiescence. This detection agrees with optical data from \kepler\ observations.
}
% conclusions heading (optional), leave it empty if necessary 
{The X-ray nature of the break frequency supports the innermost parts of the disc as source of the variability. Moreover, a similar frequency was observed in several other cataclysmic variables and a sandwich model where a geometrically thick corona surrounds the geometrically thin disc is a possible accretion configuration.}


\maketitle

\section{Introduction}
\label{introduction}

V1504\,Cyg is a dwarf nova (DN) that is located in a field intensely monitored by \kepler\, making it an ideal target for detailed variability studies. DNe are a subclass of cataclysmic variables (CVs), which are semi-detached binaries with the secondary main sequence star transferring matter toward a central white dwarf (WD) via Roche lobe overflow (see e.g. \citealt{warner1995} for a review). In the absence of strong magnetic fields an accretion disc is formed around the compact object. The disc is the dominant source of radiation in DNe. Changes in the accretion rate will thus lead to changes in the observable radiation intensity, and DNe are observed with frequent outbursts of emission.

V1504\,Cyg is a SU\,UMa system member (\citealt{raykov1988}, \citealt{nogami1997}, \citealt{cannizzo2012}) showing larger superoutbursts in addition to the regular outbursts. The latter are a long-term radiation pattern typical for all subclasses of DNe, and are governed by viscous-thermal instabilities (see \citealt{osaki1974} for the original idea) which are implemented as the disc instability model (DIM, see \citealt{lasota2001} for a review). Following this model the matter in the disc regularly alternates between two stages, i.e. a hot/ionised and a cold/neutral stage. The hot state generates outburst and is characterized by a high mass accretion rate ($\dot{m}_{\rm acc}$), while during quiescence the matter is in the cold state and $\dot{m}_{\rm acc}$ is lower. The physical principle of the DN cycle is based on the hydrogen ionisation temperature of $\sim 8000$\,K. If the mass transfer rate from the secondary is within a specific interval, $\dot{m}_{\rm acc}$ alternates between high and low values. If the mass transfer rate is high enough, the disc is in a stable hot ionised state similar to DNe in outburst. This is the case in nova like systems. Multiwavelength observations over DN cycles revealed delays between optical, UV radiation and X-rays (see e.g. \citealt{schreiber2003}). This is explained by inner disc truncation: During an outburst the disc is developed almost to the WD surface, while during quiescence the optically thick disc is truncated.

This binary has an orbital period of 100\,min (\citealt{thorstensen1997}). As typical for SU\,UMa systems, superhumps are observed during superoutbursts with a period of 104\,min (\citealt{kato2012}). Regular outbursts and superoutbursts have mean durations of 1.1-2.9 and 12\,days, respectively, while the mean inter-outburst timescale is in the larger interval of 3.2-20.4\,days (\citealt{cannizzo2012}).

%The physical origins of the superoutbursts is not the same as in regular outbursts. Two different physical scenarios have been proposed as explanations of the superoutburst, i.e. enhanced mass transfer or tidal thermal instability (\citealt{schreiber2004}). The former scenario suggests that the superoutbursts are generated when the disc mass exceeds a critical value, while the latter is based on the outer disc radius expanding to a certain critical radius (3:1 resonance radius) where the tidal activity triggers the superoutburst. \citet{osaki2013} and \citet{osaki2014} analysed \kepler\ data of V1504\,Cyg, and studied the appearance of superhumps (a variability connected to superoutbursts). The authors concluded that the superoutburst was initiated by a tidal instability (as evidenced by the growing superhump).

The short-term or fast stochastic variability (the so-called flickering) provides a powerful diagnostic tool to study the underlying accretion process. This variability has several distinct observational characteristics. The most important example is the linear correlation between the variability amplitude and log-normally distributed flux (rms-flux relation) found not only in CVs but also in symbiotic systems, X-ray binaries and active galactic nuclei (AGNs, \citealt{uttley2005}, \citealt{scaringi2012a}, \citealt{zamanov2015}). This observational phenomenon was studied and confirmed in V1504\,Cyg by \citet{vandesande2015} and \citet{dobrotka2015}. The linearity represents the multiplicative nature of variability patterns which is a basic condition for propagating accretion fluctuations as a promising model explaining flickering activity (see e.g.\citealt{lyubarskii1997}, \citealt{kotov2001}, \citealt{king2004}, \citealt{zdziarski2005}, \citealt{arevalo2006}, \citealt{ingram2010}, \citealt{kelly2011}, \citealt{ingram2013}, \citealt{cowperthwaite2014}, \citealt{hogg2016}).

Another essential observational characteristic is that flickering is represented by red noise in the power density spectra (PDS). Usually such a PDS is not a simple power law describing pure red noise, but some characteristic break frequencies or Lorentzian patterns are present (see e.g. \citealt{miyamoto1992}, \citealt{vikhlinin1994}, \citealt{sunyaev2000}). It appears that such multicomponent PDS is very common in CVs like in X-ray binaries. For such studies, a high-quality and long cadence light curve is needed. This condition is fulfilled with the long light curves obtained with the \kepler\ satellite. The first well-studied CV system using \kepler\ data was MV\,Lyr, where \citet{scaringi2012a} found four PDS components. Subsequent \xmm\ observations (\citealt{dobrotka2017}) confirmed the sandwich model interpretation (\citealt{scaringi2014}). In this model a geometrically thin but optically thick disc is surrounded by a geometrically thick optically thin disc at the inner parts. The geometrically thick disc behaves like a hot corona radiating in hard X-rays. Two other CV systems studied in detail using \kepler\ data are V1504\,Cyg and V344\,Lyr. \citet{dobrotka2015} and \cite{dobrotka2016} used these data to search for multiple component PDSs, and showed that both systems have several PDS components depending on DN stage of activity. A break frequency close to log($f$/Hz) = -3.4 is present during both, outburst and quiescence, while additional break frequencies close to log($f$/Hz) = -3 and -2.8 arise only during outburst. These characteristic frequencies present during both activity stages are interpreted as coming from the persistent part of the geometrically thin and optically thick disc, while the ones seen only during the outburst can be generated by a reformed inner disc.

Usually half of the accretion luminosity originates from the disc, while the other half is generated in the boundary layer (see e.g. \citealt{pringle1981}). If the mass accretion rate is low (< $10^{-(9-9.5)}$\,M$_{\rm \odot}$/yr$^{-1}$), the boundary layer is expected to be optically thin (\citealt{narayan1993}) and radiates mostly in hard X-rays. If the mass accretion rate is high (> $10^{-(9-9.5)}$\,M$_{\rm \odot}$/yr$^{-1}$), the boundary layer should be optically thick (\citealt{popham1995}). Such boundary layer emits EUV and soft X-rays.

In general, DNe in quiescence have low mass accretion rates, and the emitted hard X-ray spectra are well described by multitemperature collisionally ionised plasma in equilibrium with temperatures of 6-55\,keV (see \citealt{balman2020} for a review). This hard X-ray emission persists during outburst with lower fluxes and temperatures compared to quiescence. Since the mass accretion rate is high during outburst, the boundary layer is expected to be optically thick manifesting itself as black body in EUV and soft X-rays with temperatures of 5-30\,eV. Only few systems show this soft component (see e.g. \citealt{mauche1995}, \citealt{long1996}, \citealt{mauche2000}, \citealt{byckling2009}). Such search for soft components representing optically thick boundary layers in several nova-like systems in a high state were performed by \citet{balman2014}. This subclass of CVs resembles outbursting DNe, therefore the mass accretion rate should be high. The authors did not find the soft components with expected temperatures, only upper limits typical for WDs were derived using \rosat\ observations. Apparently, the presence of the soft component in CV during the high state is not common as expected, but is rather rare.

%The sandwich model proposed by \citet{scaringi2014} has important consequences. \citet{balman2014} pointed out a dilemma in mass accretion rates derived from multiwavelength observations of nova-like systems. The accretion rates derived from optical and UV luminosities resemble those of DNe in outburst, while values derived from X-ray analysis during the same brightness state resemble those in quiescent DNe. Following \citet{dobrotka2017} the sandwich model with the corona offers an explanation. Based on this interpretation, the standard geometrically thin disc with an optically thick boundary layer resembles DNe in outburst, while the surrounding corona with an optically thin boundary layer resembles quiescent DNe. However, as already mentioned, such optically thick boundary layer was not confirmed in MV\,Lyr (\citealt{balman2014}).

The sandwich model interpretation is attractive as it explains the variability in X-rays and UV. Following \citet{scaringi2014} the optical flickering in MV\,Lyr comes from the hot corona where X-ray variability is generated. These X-rays are subsequently reprocessed into optical radiation by the underlying geometrically thin disc. However, this model is energetically highly inefficient (\citealt{dobrotka2020}). The X-ray vs optical luminosity ratio is of the order of 0.001 - 0.01 as a result of the advective hot flows (ADAF) not radiating efficiently (\citealt{balman2014}, \citealt{balman2020}, \citealt{balman2022}). This means that there are not enough X-rays to explain the observed optical flickering by reprocessing, and it must be intrinsic to the optically thick disc. Moreover, \citet{dobrotka2020} studied the shot profile of the fast variability, and found that the variability has two components with similar time scale but different amplitudes. The smaller amplitude component could be generated by the reprocessing, while the dominant high amplitude central spike should come from the optically thick disc itself.

%Moreover, it is believed that the X-rays are generated in the boundary layer irrespective of the brightness state (see e.g. \citealt{balman2011}). Apparently, it is not trivial to conclude whether the boundary layer or the evaporated corona is the source of X-ray radiation, or whether the evaporated corona really exists. Such corona is well known from X-ray binaries or AGNs, and the precise geometry is still uncertain.

In this work we analyse our \xmm\ data of V1504\,Cyg with duration of 97\,ks. We base our study on spectral (Section~\ref{section_spectra}) and timing analysis (Section~\ref{flickering}). We discuss the finding in the context of the sandwich model (Section~\ref{section_discussion}). 

\section{Data}

V1504\,Cyg was serendipitously observed in the field of a short \xmm\ observation (ObsID 0743460201). It was just barely in the field of view of the two MOS cameras but not in the pn field. The data yield too low statistics and do not allow detections of the variability patterns we are interested in (\citealt{dobrotka2015}). Therefore, we have proposed a long pointed \xmm\ observation which was performed on 2017 September 26-27 for 97\,ks observation duration under ObsID 0801100101. The European Photon Imaging (EPIC) cameras MOS and pn were operated in Full Frame mode with the thin filter, and the RGS in standard spectroscopy mode. The Optical Monitor (OM) took 19 exposures in Image+Fast mode in the UVW1 filter. The data were downloaded from the XMM-Newton Science Archive (XSA), and we obtained light curves with the Science Analysis Software (SAS), version 18.0.

\begin{figure}
\resizebox{\hsize}{!}{\includegraphics[angle=-90]{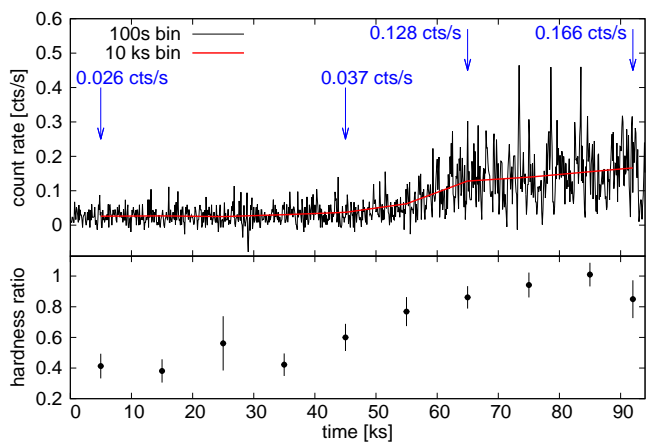}}
\caption{Upper panel - pn light curve binned into 100\,s bins (black solid line) with the long-term flux evolution shown as low resolution version of the light curve binned into 10\,ks bins (red line). The arrows with labels show the corresponding count rate (in the red line) at selected points where the trend of the low resolution light curve changes. Lower panel - hardness ratio (1.0 - 10.0\,keV band divided by 0.2 - 1.0\,keV band) calculated from low resolution light curves binned into 10-ks bins}
\label{lc_pn}
\end{figure}

For the optical monitor (OM) light curves extraction we used the standard tool {\tt omfchain} with standard extraction regions. The pn light curves were extracted from a circular region with a radius of 20\arcsec\ centred on the source, while the background was extracted from a region offset and radius of 30\arcsec. We used the {\tt epproc} tool to re-generate calibrated events files, and {\tt evselect} for light curve construction. The same we applied for MOS1 and MOS2 light curves using {\tt emproc} and {\tt evselect} tools. The pn and OM light curves are shown in Figs.~\ref{lc_pn} and \ref{lc_om}, respectively. For timing analysis we used 100\,s time bin and combined pn, MOS1 and MOS2 light curves. For the X-ray hardness ratio (lower panel of Fig.~\ref{lc_pn}) we extracted soft and hard pn light curves with splitting the energy at 1\,keV.

Finally, because of the low brightness of the object (average net pn count rate of 0.0365) the high-resolution RGS spectra are not suitable for quantitative spectral analysis even if the object is detected. Therefore, we need to rely on the low-resolution spectra from the EPIC pn detector. We extracted the pn spectrum with {\tt evselect} using the same (re-generated) calibrated pn events file used for light curve extraction. The same extraction regions as in pn light curve were used. Standardly used (recommended) quality flags were used. The rmf and arf files were generated using {\tt rmfgen} and {\tt arfgen} tools. The spectra were grouped using the {\tt specgroup} tool and rebinned to contain 25 photons per bin. Two spectra were extracted based on the brightness evolution (see next section).

\section{Light curve evolution}

\begin{figure}
\resizebox{\hsize}{!}{\includegraphics[angle=-90]{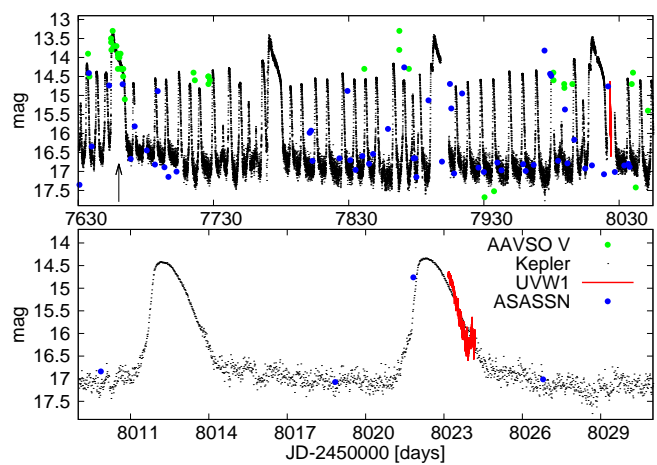}}
\caption{OM light curve (red solid line) compared to AAVSO and ASAS-SN V data and Kepler magnitudes. In the upper panel the \kepler\ light curve is shifted in time by +1775 days in order to superimpose the superoutburst in \kepler\ and AAVSO data marked by the arrow. In the lower panel the \kepler\ data are shifted horizontally by +1800.8 days to compare the OM and ASAS-SN light curve with a randomly selected regular outburst.}
\label{lc_om}
\end{figure}

The mean flux and hardness ratio of the X-ray light curve are shown in Fig~\ref{lc_pn}. There is a visible change in behaviour between approximately 50 and 60\,ks of elapsed observing time. To understand this phenomenon we compare the OM light curve with data taken from the AAVSO, ASAS-SN and \kepler\ (the same data as in \citealt{dobrotka2015}) in Fig.~\ref{lc_om}. In order to identify the activity stage during the \xmm\ observation, direct comparison of \kepler\ and \xmm\ light curves would be the best. However, we have no simultaneous \kepler\ data, and the existing \kepler\ light curve can be used at least for comparison of the phenomenology. Therefore, we shifted the \kepler\ data in time in order to "synchronise" them with one randomly selected superoutburst seen in AAVSO data (marked by the arrow in the upper panel of Fig.~\ref{lc_om}). The upper panel of Fig.~\ref{lc_om} clearly shows that the \xmm\ observation was not taken during optical quiescence. The lower panel of Fig.~\ref{lc_om} compares the OM data with a randomly selected and horizontally shifted regular outburst in the \kepler\ data. The OM and ASAS-SN data agree well with a regular outburst behaviour. Therefore, we conclude that the \xmm\ observation was taken during the decline from an outburst.

For further analysis, we divided all light curves into two subsamples, an initial (first 45\,ks in pn light curve) low-flux and a final (after 64.98\,ks in pn light curve) high-flux portion (see Fig.~\ref{lc_pn}). The region in between is excluded as a transition region. These time intervals are selected based on the low resolution light curve. They represent points where the flux changed considerably in behaviour (second and third arrow in Fig.~\ref{lc_pn}).

%Between 29 and 29.8\,ks of elapsed time, the pn light curve shows anomalous interval with count rate dropping significantly below zero suggesting the presence of non-desired events. We exclude this time interval from data analysis.

\section{Spectral study}
\label{section_spectra}

\begin{figure}
\resizebox{\hsize}{!}{\includegraphics[angle=-90]{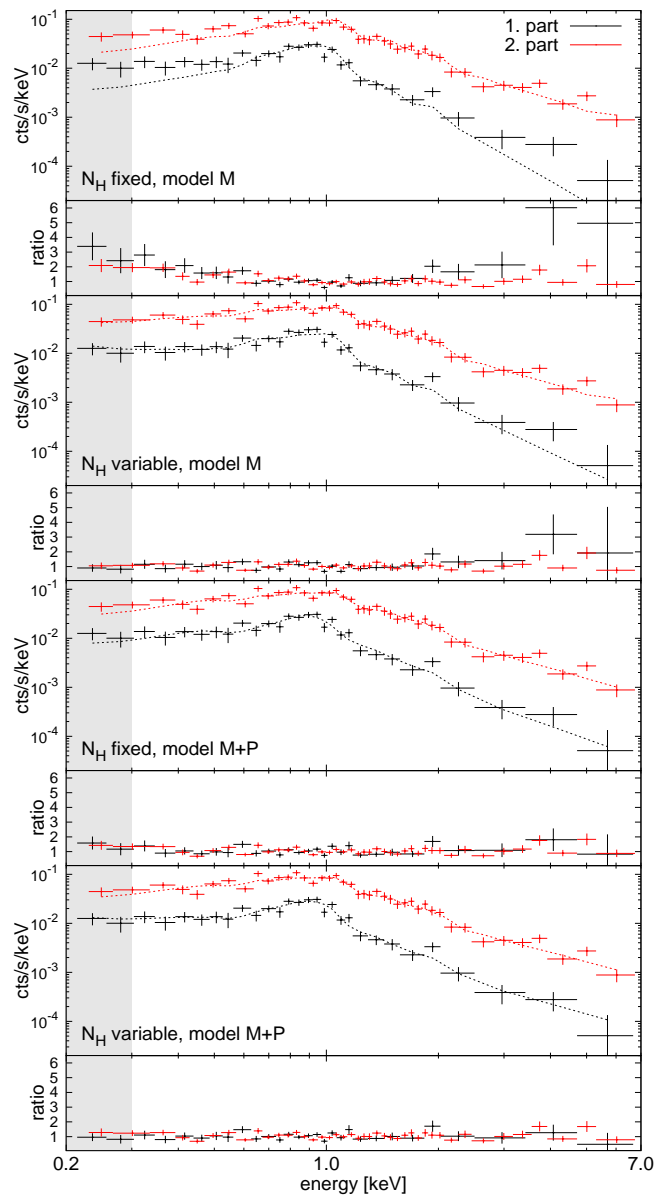}}
	\caption{EPIC pn spectra from different parts of the pn light curve with corresponding fits and ratios between data and fitted model. Parts 1 and 2 represent low and high flux portion of the data, respectively. Two models are shown (see text or Table~\ref{spectra_param} for details). The fits are performed from 0.3\,keV and the shaded area represents an extrapolated region.}
\label{spectra}
\end{figure}

We use {\tt XSPEC}\footnote{https://heasarc.gsfc.nasa.gov/xanadu/xspec/} (\citealt{arnaud1996}) to test spectral models against pn spectra via $\chi^2$ statistic. All photons below 0.3\,keV and above 10\,keV were ignored. This results in spectral bins up to approximately 7-8\,keV. Therefore, we fixed the fitted interval to 0.3 - 7.0\,keV. Since we are interested in the soft part we display the spectra in Fig.~\ref{spectra} from 0.2\,keV as allowed by \xmm\ thread\footnote{\url{https://www.cosmos.esa.int/web/xmm-newton/sas-thread-xspec}}. The models are extrapolated to this lower energy value in order to search for any soft excess. Finally, we fitted the data also from 0.2\,keV and the results are very consistent with the 0.3\,keV version described below.

We take photoelectric absorption within the neutral interstellar medium plus the photoelectric absorption of the circumstellar material into account using the {\tt tbabs} model (\citealt{wilms2000}). The only parameter of {\tt tbabs} is the neutral hydrogen column density $N_{\rm H}$ while the column density of other neutral elements in the line of sight are computed assuming cosmic abundances. As an independent estimate for the amount of interstellar absorption in the direction of V1504\,Cyg we use the HEASARC NH tool\footnote{http://heasarc.gsfc.nasa.gov/cgi-bin/Tools/w3nh/w3nh.pl}. Several (but very similar) values are computed from the HI4PI Survey (\citealt{hi4pi2016}) resulting in a mean value of $9.81\times 10^{20}$\,cm$^{-2}$.

For spectral fitting we used the cooling flow model {\tt mkcflow} (\citealt{mushotzky1988}) that \cite{mukai2003} have applied to CV spectra. The model is based on the radiative energy release in the form of optically thin plasma in local collisional equilibrium that cools in a steady-state flow. The observed radiation can be modeled with the isothermal {\tt APEC} model (\citealt{smith2001}) or a combination of them, and the {\tt XSPEC} model {\tt mkcflow} assumes the flow as an interpolation between a minimum ($T_{\rm low}$) and a maximum ($T_{\rm high}$) {\tt APEC} X-ray temperature. The {\tt APEC} model basically assumes a collisional plasma in equilibrium, thus collisional ionisation and excitations being balanced by radiative recombination and de-excitations. Such spectrum consists of bremsstrahlung continuum and emission lines. The key parameter is the electron temperature $T$ which originates from a Maxwellian velocity distribution of electrons and ions as a result of collisional equilibrium. An isothermal plasma is unlikely in nature, more likely is a broad distribution of temperatures, and the {\tt mkcflow} model is one possible realisation of a non-isothermal plasma.

The {\tt mkcflow} model has originally been developed for Galaxy Clusters for which the redshift $z$ is an important parameter. In addition to shifting the spectra, $z$ is used to obtain the flux from the model via distance/$z$. As a consequence the computation of $\dot{m}_{\rm acc}$ from the normalisation fails for $z=0$. For Galactic sources without the need for a spectral shift, we fix $z$ at a small but non-zero value, $z=5\times 10^{-8}$ like in \citet{dobrotka2017}. This is equivalent to a Doppler shift of 0.015 km\,s$^{-1}$, which is negligible.

Another free parameter is the abundance of elements heavier than hydrogen relative to Solar. In the absence of any hydrogen lines, absolute (i.e. relative to H) abundances can only be estimated assuming the number of free electrons forming the bremsstrahlung continuum is equivalent to the number of hydrogen ions. This is highly uncertain because the determination of the bremsstrahlung continuum assumes all additional emission lines are correctly included in the atomic database while a large number of weak emission lines are poorly known. Given the high uncertainty, we performed the subsequent fits with abundances fixed to solar. Anyhow, using the abundance as a free parameter does not improve the fits.

The fitted {\tt mkcflow} models are depicted in the upper two panels of Fig.~\ref{spectra}. We used the above mentioned $N_{\rm H}$ value as fixed (top) and as start value while keeping it variable (second panel). The survey value comes from interpolation of several measurements, therefore some deviation from the derived value is acceptable. Apparently, the model with fixed $N_{\rm H}$ leaves a deficiency in the soft part of both spectra which could indicate overestimated absorption or a soft excess. This is supported by the model with variable $N_{\rm H}$ yielding better agreement with the soft part of the observed spectrum. A deficit of the fit in the hard band of the first part of the spectra can also be seen. Variable $N_{\rm H}$ yields an improvement also in this case, but some hard excess is still possible. The fit parameters are listed in Table~\ref{spectra_param} as M model.
\begin{table*}
\caption{Summary of fitted parameters. We present only models tested against the pn spectrum with $\chi^2_{\rm red} < 1.5$. Parameter intervals represent the 90\% statistical uncertainties derived by {\small XSPEC}; for $T_{\rm low}$ only upper limits were found. The first and second line in each block represent the first low-flux and second high-flux portions of the pn light curve, respectively. The label M marks the results from the {\tt mkcflow} model, and the label P marks the {\tt powerlaw} model. The goodness of the fit is represented by $\chi^2_{\rm red}$ with degree of freedom $dof$. Absorbed ($\psi_{\rm abs}$) and un-absorbed fluxes ($\psi$) are in the range of $0.3 - 10.0$\,keV.}
%\caption{Summary of fitted parameters. We present only models tested against the pn spectrum with $\chi^2_{\rm red} < 2$. Parameter intervals represent the 90\% statistical uncertainties derived by {\small XSPEC}; for $T_{\rm low}$ only upper limits were found. The first and second line in each block represent the first low-flux and second high-flux portions of the pn light curve, respectively. The label M marks the results from the {\tt mkcflow} model, and the label P marks the {\tt powerlaw} model. The goodness of the fit is represented by $\chi^2_{\rm red}$ with degree of freedom $dof$. Absorbed ($\psi_{\rm abs}$) and un-absorbed fluxes ($\psi$) are in the range of $0.2 - 15.0$\,keV.}
\begin{center}
\begin{tabular}{ccccccccc}
\hline
\hline
model & $N_{\rm H}$ & $T_{\rm low}$ & $T_{\rm high}$ & $\dot{m}_{\rm acc}/10^{-12}$ & ph. index & $\chi^2_{\rm red}/dof$ & $\psi_{\rm abs}/10^{-13}$ & $\psi/10^{-13}$\\
$\times {\tt tbabs}$ & $10^{20}$\,cm$^{-2}$ & (keV) & (keV) & (M$_{\rm \odot}$/yr) & & & (erg/cm$^2$/s) & (erg/cm$^2$/s)\\
\hline
M & $< 2.08$ & $< 0.27$ & 1.79 - 2.64 & 0.47 - 0.70 & & 1.46/23 & 0.37 & 0.37\\
& 1.79 - 5.67 & $< 0.41$ & 6.12 - 8.72 & 1.02 - 1.45 & & 1.38/36 & 2.97 & 3.22\\
%M & $< 1.94$ & $< 0.27$ & 1.79 - 2.63 & 0.47 - 0.70 & & 1.31/26 & 0.39 & 0.40\\
%& 2.02 - 4.82 & $< 0.42$ & 6.36 - 9.18 & 1.01 - 1.41 & & 1.31/38 & 3.15 & 3.49\\
\hline
M+P & 9.81 & $< 0.76$ & 0.66 - 1.41 & 0.73 - 2.08 & 2.09 - 2.90 & 1.18/22 & 0.41 & 0.61\\
& 9.81 & $< 0.39$ & 1.62 - 4.30 & 1.12 - 1.90 & 1.81 - 2.10 & 1.36/35 & 2.94 & 3.77\\
%M+P & 9.81 & $< 0.77$ & 0.66 - 1.41 & 0.73 - 1.86 & 2.39 - 2.96 & 1.11/25 & 0.42 & 0.77\\
%& 9.81 & $< 0.42$ & 1.61 - 4.63 & 1.04 - 1.76 & 1.88 - 2.18 & 1.39/37 & 3.25 & 4.48\\
\hline
M+P & $< 10.29$ & $< 0.78$ & 0.60 - 1.73 & 0.56 - 1.02 & 1.57 - 2.67 & 1.12/21 & 0.45 & 0.49\\
& 5.12 - 12.31 & $< 0.42$ & 2.22 - 6.48 & 1.10 - 1.89 & 1.43 - 2.10 & 1.37/34 & 3.05 & 3.62\\
%M+P & $< 8.72$ & $< 0.78$ & 0.70 - 1.74 & 0.58 - 1.07 & 1.44 - 2.61 & 1.01/24 & 0.48 & 0.56\\
%& 2.15 - 8.48 & $< 0.43$ & 2.12 - 7.90 & 1.08 - 1.67 & 0.36 - 2.05 & 1.32/36 & 3.50 & 4.03\\
\hline
\end{tabular}
\end{center}
\label{spectra_param}
\end{table*}

We tested an alternative scenario to improve the soft deficit in the model with fixed $N_{\rm H}$. \cite{balman2014} added a {\tt powerlaw} model to their spectral fits to \swift\ spectra of MV\,Lyr and interpreted it as possible scattering effects of X-rays from a wind or extended component. Resulting fits with fixed or variable $N_{\rm H}$ are depicted in Fig.~\ref{spectra} as M+P models (parameters summarised in Table~\ref{spectra_param}). Apparently, adding this component to the {\tt mkcflow} can also remedy the soft deficit, and it improves also the hard band fit of the first part of the spectra. While this improvement is significant in the first part of the spectra where the $\chi^2_{\rm red}$ decreased from 1.46 to 1.12, the second part did not improve. This improvement of the fits using an additional {\tt powerlaw} component only in the first part and not in the second is supported using F-test.

All models yield upper limits of $T_{\rm low}$ below 1\,keV, and $T_{\rm high}$ tends to be lower in the first part of the spectra. Moreover, the latter parameter reaches considerably lower values when a {\tt powerlaw} model is added. The $\dot{m}_{\rm acc}$ parameter tends to be higher in the spectrum extracted from the second episode of the light curve\footnote{This is not valid for the values derived from M+P model with fixed $N_{\rm H}$ parameter, but based on best fits it is valid.}. Since $\dot{m}_{\rm acc}$ is the normalisation of the cooling flow model, the rise of this parameter is a result of increasing luminosity during the second part of the light curve. The physical reason of such luminosity increase can be higher radiation efficiency, therefore the $\dot{m}_{\rm acc}$ values must be taken with caution.

%Finally, we used the same value of $N_{\rm H}$ for both first and second part of the data\footnote{We fitted both spectra simultaneously.}. However, when trying different values by fitting every part of the spectrum independently, $N_{\rm H}$ did not increase above the survey derived value for any of the spectra which would suggest additional circumstellar absorption.

Finally, we tried also to add a black body component in order to investigate possible soft excess attributed to an optically thick boundary layer. We did not get any fit improvement. The resulting $\chi^2_{\rm red}$ remained almost the same or little worse compared to models without black body. Moreover, extrapolated models in Fig.~\ref{spectra} down to 0.2\,keV do not show any noticeable soft excess. Therefore, we conclude that no black body component can be identified.

\section{Fast variability study}
\label{flickering}

\citet{dobrotka2015} and \citet{dobrotka2016} studied optical flickering of V1504\,Cyg finding a break frequency at log($f$/Hz) = -3 during the outburst. They proposed that this characteristic break frequency can originate from the inner parts of the geometrically thin disc which was truncated during quiescence. However, a very similar value was found in optical (\citealt{scaringi2012a}) and X-ray (\citealt{dobrotka2017}) data of MV\,Lyr. The presence in both bands imply that the sandwich corona can be the origin of the variability. MV\,Lyr shows also potential variability with higher frequency of log($f$/Hz) = -2.4 found in \xmm\ (\citealt{dobrotka2019}) and \nustar\ light curves (\citealt{balman2022}), although the latter having large error interval in the range from -2.2 to -2.9. For the same reason we search for any variability pattern in our X-ray data of V1504\,Cyg.

\subsection{Power density spectrum study}

For the PDS study we divided the light curve into several subsamples. Subsequently, using the Lomb-Scargle algorithm\footnote{We used python's package {\tt Astropy} (\citealt{astropy_collaboration2013,astropy_collaboration2018,astropy_collaboration2022}).} \citep{scargle1982} we calculated log-log periodograms for each of these subsamples. These periodograms were averaged and re-binned into equally spaced bins with a minimum number of periodogram points per bin. The averaging is performed over log($p$) rather than $p$ following \citet{papadakis1993}. Moreover, such averaging of log($p$) yields symmetric errors (see e.g. \citealt{vanderklis1989}, \citealt{aranzana2018}). The averaging and binning reduce the PDS noise, but affect the frequency resolution of the PDS. Therefore, an empirical compromise between noise and resolution must be found. The PDS low frequency end is defined by the duration of each light curve subsample while the high frequency end is determined by the Nyquist frequency. Resulting PDSs were rms normalised (\citealt{miyamoto1991}).

All PDSs estimated in this work are calculated using light curves with 100\,s resolution splitted into three subsamples, and re-binned using bin width of 0.1 in log scale, and a minimum number of points per bin of $3 \times 3$ (3 subsamples, and 3 points from each corresponding periodogram).

As confidence test we used simulations following \citet{timmer1995}. For this purpose the observed PDSs were fitted with a broken power law fit. In the log-log representation such fit consists of a decreasing linear function with frequency describing the red noise, and a constant for highest frequencies as Poisson noise if needed. The red noise slope summarised in Table~\ref{pds_fits} is used as input parameter for the simulations, and red noise light curves with added Poisson noise are calculated. The simulated light curves have the same sampling, mean count rate and amplitude of variability $\sigma_{\rm rms}$ (see next section) as observed data. If a long-term trend is present (OM data), the light curve is first detrended using a polynomial. The red noise slope for simulations is measured after the detrending, and the polynomial is added to every simulated light curve. Subsequently, a PDS is calculated using the same method and parameters setup as used for the real data. We repeated the process 1000 times, and the mean value of the power with $\sigma$ is calculated. For detection of significant patterns we used at least 2-$\sigma$ level.
\begin{table}
\caption{Red noise slopes of final PDS models, i.e. broken power law with break frequency in the second part of the EPIC light curve, and simple red noises in other subsamples.}
\begin{center}
\begin{tabular}{cccc}
\hline
\hline
subsample & EPIC & OM\\
\hline
First part & -- & $-0.65 \pm 0.16$\\
Second part & $-1.51 \pm 0.48$ & $-0.84 \pm 0.16$\\
\hline
\end{tabular}
\end{center}
\label{pds_fits}
\end{table}

\begin{figure}
\resizebox{\hsize}{!}{\includegraphics[angle=-90]{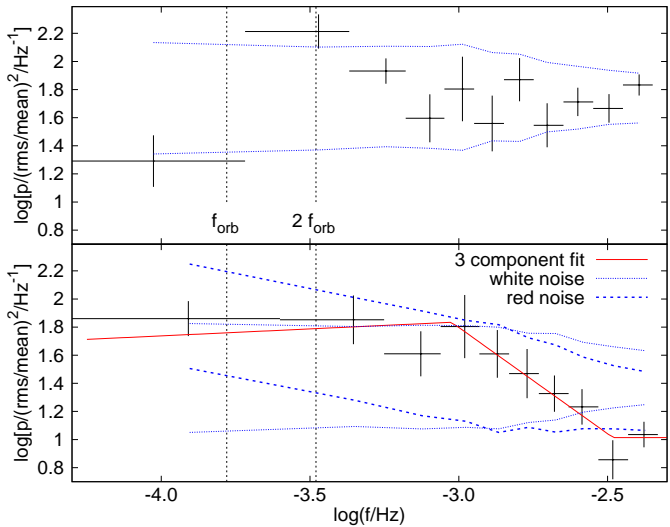}}
\caption{PDSs calculated from combined EPIC light curves from first (upper panel) and second (lower panel) part of the data. Crosses are the mean values of the PDS power at a mean frequency of every bin. The error bars represent bin widths and errors of the mean. Vertical dotted lines show the orbital frequency $f_{\rm orb}$ and its first harmonics. Blue lines represent the 2-$\sigma$ confidence intervals from simulations of noises.}
\label{pds_epic}
\end{figure}

As input X-ray light curves we used combined pn, MOS1 and MOS2 light curves, because using only the pn data yield too low confidence of PDS patterns. Fig.~\ref{pds_epic} shows the resulting PDSs from the first and second part of the data with durations of 40\,ks and 30.5\,ks, respectively. All points in the first part lies within 2-$\sigma$ limit of a white noise. Therefore, we conclude that PDS of the first part is flat and dominated by Poisson noise. The second part of the data already shows some trend. There is a break at log($f$/Hz) = $-3.02 \pm 0.13$. Above this frequency, the power decreases as expected from red noise, and the Poisson white noise level dominates from log($f$/Hz) = $-2.48 \pm 0.11$. Apparently, the white noise 2-$\sigma$ interval is not able to reproduce the PDS. Using the red noise model instead improves the situation, but the PDS point at log($f$/Hz) = -2.48 still lies below the 2-$\sigma$ limit. In order to describe all PDS points, a multicomponent model depicted by the solid red line in Fig.~\ref{pds_epic} is needed. Therefore, the break at log($f$/Hz) = -3.02 is real with approximately 2-$\sigma$ confidence.

\begin{figure}
\resizebox{\hsize}{!}{\includegraphics[angle=-90]{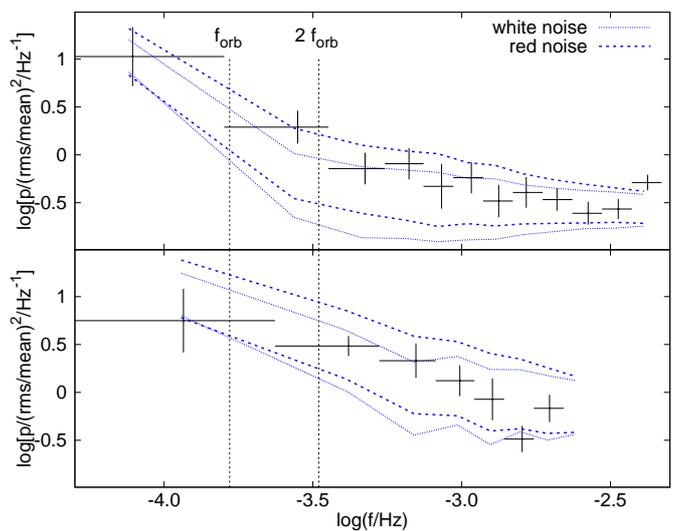}}
\caption{The same as Fig.~\ref{pds_epic} but for OM data.}
\label{pds_om}
\end{figure}

Fig.~\ref{pds_om} shows PDSs calculated from first and second part of OM data with durations of 49.3\,ks and 23.2\,ks, respectively. These light curves show strong long-term trend which was accounted for during the simulations. Upper panel of Fig.~\ref{pds_om} shows 2-$\sigma$ intervals for the first part of the OM data. The white noise case does not describe the PDS well, while the red noise model is much better and all PDS points are within the 2-$\sigma$ interval. Therefore, the red noise nature of the variability is clear. The lower panel of Fig.~\ref{pds_om} shows PDS calculated from second part of the OM data. Investigation of the variability nature is more difficult because of short subsample duration. Both white and red noise cases describe all data points within 2-$\sigma$ interval, but subjectively the red noise case matches better. We conclude, that even if the red noise nature of the variability is not conclusive, it is more likely.

\subsection{Amplitude of variability}

The absolute amplitude of variability is defined as square-root of the variance, i.e.
\begin{equation}
\sigma_{\rm rms} = \sqrt{\frac{1}{N - 1} \sum^N_{i = 1} (\psi_i - \overline{\psi})^2},
\end{equation}
where $N$ is the number of the light curve points used to calculate $\sigma_{\rm rms}$, $\psi_i$ is the $i$-th flux point and $\overline{\psi}$ is the mean value of $\psi_i$ calculated from the same $N$ points. In all subsequent analysis we used $N = 10$.

\begin{figure}
\resizebox{\hsize}{!}{\includegraphics[angle=-90]{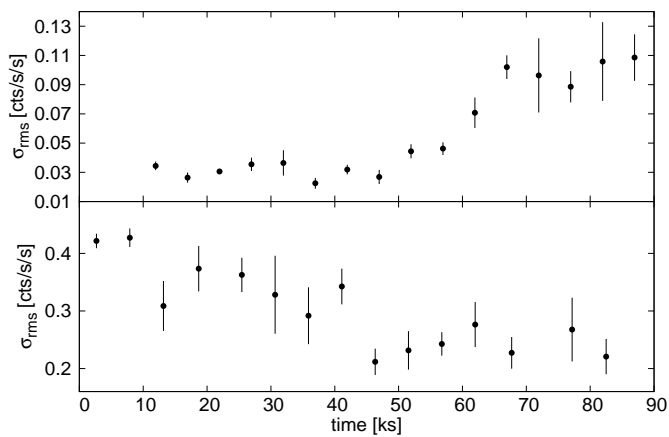}}
\caption{Amplitude of variability $\sigma_{\rm rms}$ evolution with errors of the mean as uncertainty estimate for combined EPIC (upper panel) and OM (lower panel) light curves.}
\label{rms}
\end{figure}

Fig.~\ref{rms} represents the $\sigma_{\rm rms}$ time evolution of both the pn and OM light curves. 5 adjacent $\sigma_{\rm rms}$ points are averaged to obtain the mean $\sigma_{\rm rms}$. The different time evolution in both bands is obvious. The $\sigma_{\rm rms}$ in X-rays shows two different stages. This behaviour is similar to the mean flux evolution represented by the low resolution light curve in Fig.~\ref{lc_pn}. Apparently, the X-ray $\sigma_{\rm rms}$ is correlated with the mean flux. The same is valid also for the UV light curve, but its time evolution is different. The $\sigma_{\rm rms}$ just decreases with the UV flux, and does not show any significant deviation in behaviour related to the two-phase character seen in the X-ray data.

\section{Discussion}
\label{section_discussion}

In this work we studied our observation of V1504\,Cyg obtained by the \xmm\ observatory. The observation was motivated by the previous studies of V1054\,Cyg optical data taken by the \kepler\ spacecraft (\citealt{dobrotka2015}, \citealt{dobrotka2016}).

The observation was taken during the decline from an outburst, and a transition phase was captured. This transition divides the light curve into two separate phases with different characteristics. For this reason we needed to split the observation into two subsamples. However, the transition phase is very unique and allowed us to follow reconfiguration of the accretion disc in time.

\subsection{Light curve behaviour}

The X-ray light curve in Fig.~\ref{lc_pn} shows a striking change in X-ray behaviour near 60\,ks after the start of the observation, i.e. the overall mean flux, hardness of the radiation, and the amplitude of the variability $\sigma_{\rm rms}$ increased to a higher constant level by a factor of approximately 4.7, 2.5 and 2.7, respectively. While the X-ray flux rises, the UV radiation decreases (Figs.~\ref{lc_pn} and \ref{lc_om}).

Following DIM (see \citealt{lasota2001} for a review) which describes a DN cycle, the observed X-ray behaviour is very typical for a decline from an outburst. During this decline a cooling front with a rarefaction wave propagates through the disc. As a consequence the disc is depleting the accumulated matter and the surface density with overall $\dot{m}_{\rm acc}$ decreases. The disc starts to be truncated because the inner disc evaporates by coronal siphon flow due to insufficient cooling (\citealt{meyer1994}, \citealt{meyer2000}), and an inner hot optically thin and geometrically thick disc forms. Such a hot disc is an ADAF (\citealt{narayan1994}, \citealt{narayan1995}, \citealt{abramowicz1995}) or corona radiating in X-rays. Such ablation of the inner disc via evaporation describes well the observed delay between optical emission, UV and X-rays in DNe (\citealt{hameury1999}). As a consequence of such inner hole formation, the UV radiation generated by the receding inner disc decreases, while the X-ray radiation rises in flux because of a rebuilding corona. A similar X-ray brightness increase during the decline from an outburst is seen in DIM simulations (see e.g. \citealt{schreiber2003}) and observations of DNe (see SS\,Cyg for example, \citealt{wheatley2003}, \citealt{mcgowan2004}).

Crucial is the different time evolution of $\sigma_{\rm rms}$ in UV and X-rays seen in Fig.~\ref{rms}. This different behaviour of the two bands suggests different physical origins of the radiation variability in the two bands. The localisation of the sources is clear, i.e. the UV comes from the receding inner geometrically thin disc, while the X-rays from the rebuilding corona or boundary layer.

\begin{figure}
\resizebox{\hsize}{!}{\includegraphics[angle=0]{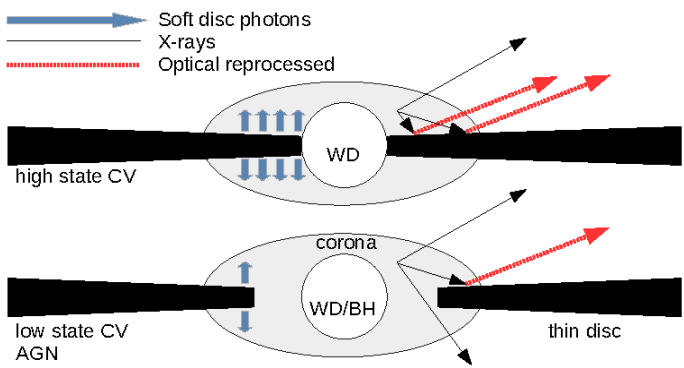}}
\caption{Illustration of a possible disc configuration for DNe in different brightness states. The corona is a complex structure containing ADAF in the low state of a CV. The thin arrows show the relation between X-rays, reprocessed UV radiation and inner geometrically thin disc as reprocessing region. Much less is reprocessed into optical when the disc is truncated. Thick blue arrows show soft disc photons as source of Compton cooling of the corona.}
\label{model}
\end{figure}

The different X-ray and UV characteristics are well known from some AGNs. \xmm\ observations of IRAS\,13224-3809 show that the UV variability is not correlated with X-rays (\citealt{buisson2018}). \citet{fabian2009} and \citet{pawar2017} showed the same also for 1H\,0707-495. \citet{pawar2017} concluded that the discrepancies between the two energy bands rule out reprocessing of X-rays as the source of UV variability. AGNs are a special case, where the inner disc is truncated, resembling DNe in quiescence. The corona radiates in X-rays, and if an inner geometrically thin disc as reprocessing region exists, the X-rays are reprocessed into optical and UV (upper illustration of Fig.~\ref{model}). As a consequence these two bands should have the same characteristics like PDS structure or $\sigma_{\rm rms}$ evolution. This is the case of DNe in their high state, but it is better demonstrated by the high state of the nova-like variable MV\,Lyr (\citealt{dobrotka2017}). If the inner geometrically thin disc is truncated (lower illustration of Fig.~\ref{model}), a large part of the reprocessing region is missing and the detected UV radiation is not generated by reprocessing. In such case the UV should be generated by another mechanism. The intrinsic variability of the accretion flow through the remaining outer disc via propagating fluctuations is a possibility. Such variability has "its own" characteristics not depending on X-rays. \citet{balman2012} reported correlations with time lags of 96-181\,s between UV and X-rays in five CVs. The authors interpret the lag as travel time of matter from a truncated inner disc to the WD surface\footnote{We can not confirm nor deny such correlation in V1504\,Cyg, because the cross-correlation function is unclear and yield no conclusive results. Its shape is more or less flat mainly in the first low flux portion of the light curve. This is not surprising because compared to for example MV\,Lyr where such crosscorrelation was studied (\citealt{dobrotka2017}) our X-ray light curve of V1504\,Cyg has 10 to 100 times lower count rate.}, therefore supporting the non-reprocessing scenario. However, some partial irradiation of the remaining disc is still possible, but the reprocessed UV power can be too weak to dominate the PDS or $\sigma_{\rm rms}$ characteristics of the light curve. Such reprocessing in quiescent DNe was shown by \citet{balman2012}, \citet{aranzana2018} and \citet{nishino2022} using time lag analysis.

\subsection{Power density spectra}

While we see no break frequency nor evolution in OM PDSs, the situation is different in X-rays. The first part of the observation with lower flux is flat, i.e. does not show any characteristic frequency, nor red noise. The second part of the data with higher flux is characterised by a PDS showing significant red noise, and a possible break at log($f$/Hz) = -3.02.

This behaviour is very similar to SU\,UMa (\citealt{balman2015}). The authors interpret the break as finger print of the inner truncated disc. The radius of the truncated disc in quiescence is further away from the WD, and it is closer to the WD during outburst. As a consequence, the Keplerian frequency should be lower in the former case. Moreover, if the disc reaches almost the surface of the WD, the break can disappear. The same radius vs frequency interpretation present \citet{balman2012} in several other DNe. Following the authors SS\,Cyg in outburst does show break in X-rays. Moreover, \citet{dobrotka2016} showed different behaviour for optical PDSs of V1504\,Cyg and V344\,Lyr. Both systems show dominant frequency close to log($f$/Hz) = -3 during outburst, which disappear in quiescence. Apparently, the disappearance of break frequencies in PDSs during outburst is not always present. Important is to take into account the exact timing of the observation. If the break frequency is connected to the inner disc truncation, this truncation is evolving from outer to inner radii during the rise of the outburst and reverse back during the outburst decline. If the observation is taken during the outburst but closer to the rising or declining part, the truncation as a break frequency could still be detected. However, \citet{dobrotka2016} took into account this timing and used various intervals covering the outburst. All cases showed the same behaviour, i.e. optical PDSs show dominant frequency close to log($f$/Hz) = -3 during outburst, which disappear in quiescence. On the other hand, our non detection of any PDS pattern during the first part of the V1504\,Cyg X-ray light curve can be a simple result of too low count rate.

Moreover, if the dominant break frequency in PDS is a finger print of the inner truncated radius, this frequency should increase during the transition from quiescence to outburst and vice versa. The opposite was shown by \citet{dobrotka2020} using \kepler\ data of MV\,Lyr. The latter is a nova-like CV system in a high state, but experiencing transitions to low states and back. \kepler\ observations of this binary recorded such transition. During the decline from the high state, during which the disc truncation radius should increase, the dominant frequency at log($f$/Hz) = -3 did not decrease, but contrary increased.

Even if the scenario where the dominant PDS characteristic frequency is a finger print of the truncation radius is attractive and similar to the X-ray binaries, there is an alternative. \citet{scaringi2014} explained the presence of log($f$/Hz) = -3 frequency in optical PDS of MV\,Lyr by reprocessing of X-rays into optical. The origin of these X-rays should be the geometrically thick inner corona. If the dominant PDS characteristic frequency is a finger print of for example the radial extent of such corona, and not the inner standard geometrically thin disc, the shrinking of the former during transition from high to low state can produce characteristic (Keplerian) frequency increase, rather than decrease (see top panel of Fig.~3 in \citealt{scaringi2014}). Such radial "shrinking" of corona was observed also in AGNs by \citet{kara2013} based on soft time lags in IRAS\,13224-3809, and \citet{wilkins2014} by modeling the relativistically broadened iron K$\alpha$ fluorescence line in 1H\,0707-495. The authors conclude that the X-ray source is more compact during low flux states. Moreover, both authors conclude that the spatial extent of the corona can be interpreted as an increase in the height of the corona above the disc, or/and some radial expansion as well (see e.g. Fig.~9 of \citealt{wilkins2015}). However, the AGN case is much more complex. The corona may contain sandwich and lamp-post components, and during the flux changes the corona can evolve or change between these two configurations. Anyhow, this confirms that a sandwich corona is not stable and radial variability is possible.

%The similarity between V1504\,Cyg and MV\,Lyr is based on detection of break frequency at log($f$/Hz) = -3 both in optical and X-rays during (MV\,Lyr) or close to (V1504\,Cyg) high state\footnote{Important would be to see whether the break frequency at log($f$/Hz) = -3 is seen during the quiescence in X-rays, because it disappears in optical (\citealt{dobrotka2015,dobrotka2016}).}. However, the orbital periods of these binaries are different; 1.7\,h in V1504\,Cyg (\citealt{thorstensen1997}) vs 3.3\,h in MV\,Lyr (\citealt{borisov1992}). This period is crucial because many other system parameters depend on it. If the detected frequency and its physical origin is independent on these parameters, it supports the idea that the frequency is not a finger print of the disc truncation radius. This was studied by \citet{dobrotka2020} who proposed that the characteristic frequency at log($f$/Hz) = -3 is common in various other systems in a high state, while it is absent in the low state. The hot X-ray corona as origin of the variability as proposed by \citet{scaringi2014} is an alternative. In such case the coronal flow generates $\dot{m}_{\rm acc}$ fluctuations which propagates inward. These fluctuations modulate the X-ray radiation which is reprocessed into optical by the underlying optically thick disc. If such disc is truncated, much smaller surface is left for the reprocessing (Fig.~\ref{model}).

The similarity between V1504\,Cyg and MV\,Lyr is based on detection of a break frequency at log($f$/Hz) = -3 both in optical and X-rays during high state or close to it. Important would be to see whether the break frequency at log($f$/Hz) = -3 is seen during the quiescence in X-rays, because it disappears in optical in both systems (\citealt{dobrotka2015,dobrotka2016,dobrotka2020}). \citet{dobrotka2020} proposed that the characteristic frequency at log($f$/Hz) = -3 is common in various other systems in a high state, while it is absent in the low state. The hot X-ray corona as origin of the variability as proposed by \citet{scaringi2014} is a possible model explaining such behaviour. In such case the coronal flow generates $\dot{m}_{\rm acc}$ fluctuations which propagate inward. These fluctuations modulate the X-ray radiation which is reprocessed into optical by the underlying optically thick disc. If such disc is truncated, much smaller surface is left for the reprocessing (Fig.~\ref{model}). As a consequence the optical radiation does not show the same timing characteristics as X-rays, i.e. the break frequency at log($f$/Hz) = -3 is seen in X-rays but not in UV like in our case.

However, this interpretation is in contrast with missing break frequency in the first part of the X-ray light curve of V1504\,Cyg which is closer to the outburst. There are two possibilities how to explain it. One is the solution proposed by \citet{balman2012} that the break frequency is generated by the inner disc edge in a truncated disc. Such disc reaches the WD surface during the outburst and no break is expected. However, this is in contrast with the detection of the log($f$/Hz) = -3 signal in optical \kepler\ data during the outburst (\citealt{dobrotka2015}, \citealt{dobrotka2016}). Moreover, as already mentioned the frequency evolution during transition to low state in MV\,Lyr (\citealt{dobrotka2020}) does not support the inner disc edge scenario in CVs. If still true, the optical variability at log($f$/Hz) = -3 in \kepler\ data of V1504\,Cyg should be generated by the inner region of optically thick disc, and the existence of the same break frequency at log($f$/Hz) = -3 in both optical \kepler\ and second part of the X-ray data is not explained. It can be just a coincidence. Second explanation is based on sandwich model inspired by MV\,Lyr case (\citealt{scaringi2014}). On the rising or declining branch of an outburst but still far from the maximum the inner geometrically thick corona generates X-ray variability with frequency log($f$/Hz) = -3. We detected it in the second part of X-ray light curve closer to quiescence. These X-rays are not sufficiently reprocessed into optical because of inner disc truncation (Fig.~\ref{model}), and the optical variability with frequency log($f$/Hz) = -3 is missing. During or around the outburst maximum the lower X-ray temperature implies less evaporation. As a consequence the corona becomes very weak, and the X-ray variability is hardly detectable like in the first part of our X-ray light curve closer to outburst maximum. Since the X-ray power is too weak, the reprocessing can not explain the optical variability seen in \kepler\ data. Like in the previous scenario this optical variability must be generated by the inner optically thick disc. In the outburst the gas in this disc is evaporated into the corona, but re-condensates back (\citealt{meyer2007}). If the corona generates the log($f$/Hz) = -3 variability via mass accretion fluctuations, these fluctuations re-condensate and propagate in the thin disc. This can generate optical variability with the same frequency. Finally, since we did not detect the log($f$/Hz) = -3 variability in the first part of our X-ray nor UV light curve, the corona is still weak and the optically thick disc is already slightly truncated. Therefore, the sandwich model is able to explain the connection of X-rays with optical and detection of the same frequency in both bands.

The re-condensation of the coronal flow can yield another consequence. After such re-condensation the accretion flow via geometrically thin disc is a mixture of the "original" flow with its own fluctuations, and the re-condensed contribution. Usually an accretion flow with propagating mass accretion fluctuations shows linear rms-flux relation. Therefore a mixture of two flows with different fluctuations due to different characteristics should deform the linearity of the rms-flux relation. This was observed by \citet{dobrotka2015} in \kepler\ data of V1504\,Cyg. While the quiescent light curve has the expected linearity, the outburst data show the linear relation plus another component generating offset in the flux axis. This offsetting flux contribution can be generated by the "original" flow, and the variability with linear rms-flux relation together with log($f$/Hz) = -3 frequency can be the result of the re-condensation.

%Finally, different behaviour of X-ray and UV PDSs supports the discussion from the previous section about different origin of UVs and X-rays.

\subsection{Spectral analysis}

The spectral fitting in Section~\ref{section_spectra} brought two important results worth discussing, i.e. $T_{\rm high}$ and $\dot{m}_{\rm acc}$ measurements.

In general, $T_{\rm high}$ parameter tends to be higher in the second part of the data. The final values depend on the best model. While the simpler model M is enough to describe the second part of the data, the first part of the observation is improved after adding a power law component to the fitted model (M+P). Taking the best fits (M+P model for the first part and M model for the second part of the data) the derived $T_{\rm high}$ 90\% confidence intervals are 0.6 - 1.7 and 6.1 - 8.7\,keV, respectively.

V1504\,Cyg with its best fitted value intervals is cooler closer to the outburst (first part of the data). This behaviour is expected, because typical $T_{\rm high}$ derived from X-ray spectra of DNe in outburst should be cooler compared to quiescence (see e.g. \citealt{wheatley2003}, \citealt{mcgowan2004}, \citealt{ishida2009}, \citealt{collins2010}, \citealt{balman2015}). This is well demonstrated by AAVSO vs. \rxte\ observations of the prototype DN SS\,Cyg (\citealt{mcgowan2004}). The quiescent temperature of this CV is reaching 26\,keV derived from simple bremsstrahlung model, and it decreases to about 6\,keV during optical outburst. Fitting a thermal plasma model to X-ray spectra of the same object yield temperatures of 17 and 7\,keV in quiescence and outburst, respectively (\citealt{wheatley2003}). Moreover, the outburst of a DN is similar to nova-like systems in the high state. \citet{balman2014} measured $T_{\rm high}$ parameter in few nova-likes using \swift\ observations. They derived $T_{\rm high}$ with values greater than 20\,keV. In the case of MV\,Lyr this was confirmed by \xmm\ observation by \citet{dobrotka2017}. Therefore, V1504\,Cyg in a high state is considerably cooler compared to nova-likes and prototypical SS\,Cyg in similar stage. Closer to V1504\,Cyg temperatures is SU\,UMa case with values of 8 and 4\,keV in quiescence and outburst, respectively (\citealt{collins2010}). Finally, V1504\,Cyg during outburst resembles more the CV system WZ\,Sge (\citealt{balman2015}). The fitted temperature of WZ\,Sge increased from 1\,keV during the outburst to 30\,keV in quiescence. The former match well our measurement of V1504\,Cyg, while the latter is considerably larger. However, our observation was taken on the way to quiescence, therefore the "final" physical conditions of the quiescent flow were not yet stabilised.

Moreover, during the outburst of WZ\,Sge a power law component in X-ray spectra is present which disappears toward quiescence (\citealt{balman2015}). It suggests thermal Comptonization of the optically thick disc photons and/or scattering from the wind. The same result was derived by \citet{balman2014} from X-ray spectra of BZ\,Cam and MV\,Lyr in the high state which resembles DNe in outburst. A power law component was needed to find an acceptable fit in these two nova-like CV systems, and is explained by the presence of a wind outflow. The presence of a power law component closer to the outburst maximum (first part of the observation) in our observation of V1504\,Cyg matches the mentioned cases and suggests the presence of a wind outflow. Therefore, our study of V1504\,Cyg together with other findings implies that wind outflows are common in the high states of CVs. Finally, thermal Comptonization of the soft disk photons can still be another or additional cause of the power law in outburst spectra. The proposed geometry in Fig.~\ref{model} describes why such Comptonization should be stronger in a high state. The geometrically thin disc penetrates deeper underneath the central part of the corona and the irradiation by soft photons from the optically thick disc is stronger. This results in corona cooling via inverse Compton process, and the X-ray temperature is lower. This process is known from X-ray binaries (see e.g. \citealt{done2007} for review).

The $\dot{m}_{\rm acc}$ parameter 90\% confidence intervals derived from best spectral fits are $(0.56 - 1.02) \times 10^{-12}$\,M$_{\rm \odot}$/yr and $(1.02 - 1.45) \times 10^{-12}$\,M$_{\rm \odot}$/yr for the first and second part of the data, respectively. However, this spectrally derived parameter is depending on red shift which was estimated very roughly. Another way of estimating $\dot{m}_{\rm acc}$ is to use the theoretical assumption that approximately half of the gravitational potential energy is liberated by the interaction of the matter flow and the WD; $L_{\rm X} = GM_{\rm WD}\dot{m}_{\rm acc}/2r_{\rm WD}$, where $L_{\rm X}$ is the X-ray luminosity, $G$ is the gravitational constant, $M_{\rm WD}$ is the WD mass and $r_{\rm WD}$ is the WD radius. As input parameters we use V1504\,Cyg {\it Gaia} distance of 553.5\,pc (\citealt{bailer-jones2021}), best fitted unabsorbed luminosities from Table~\ref{spectra_param} and $r_{\rm WD}$ estimate from analytical approximation of the mass-radius relation by \citet{nauenberg1972}. In order to get $\dot{m}_{\rm acc}$ in the ranges summarised in Table~\ref{spectra_param} we need a WD mass of 0.44 - 0.64\,M$_{\rm \odot}$ and 1.03 - 1.16\,M$_{\rm \odot}$ for the first and second part of the data, respectively. The resulting WD masses are realistic and the different values probably result from different radiation efficiency in the two brightness stages. Assuming a WD with a mass of 1\,M$_{\rm \odot}$ and unchanged radiation efficiency, the $\dot{m}_{\rm acc}$ values are $0.23 \times 10^{-12}$\,M$_{\rm \odot}$/yr and $1.54 \times 10^{-12}$\,M$_{\rm \odot}$/yr. Apparently, the unknown WD mass and rough estimate of the red shift brings some uncertainty in $\dot{m}_{\rm acc}$ estimate, but important is that the value of radiation efficiency dependent $\dot{m}_{\rm acc}$ is of the order of $10^{-13}$ and $10^{-12}$\,M$_{\rm \odot}$/yr as derived with the spectral modeling.

Based on the general phenomenology of DNe, the brightness decline from an outburst is generated by decreasing $\dot{m}_{\rm acc}$. Based on DIM simulations $\dot{m}_{\rm acc}$ should decrease from for example $10^{-9}$ to $10^{-14}$\,M$_{\rm \odot}$/yr (see e.g. Fig.~3 of \citealt{hameury2000}). Therefore, it seems like the rising $\dot{m}_{\rm acc}$ derived spectroscopically from our \xmm\ observation of V1504\,Cyg during the outburst decline contradicts this expectation. Nevertheless, during decline from an outburst the inner disc is truncated via evaporation (\citealt{meyer1994}) into radiatively inefficient ADAF (\citealt{meyer2000}). As a result the advective characteristics of the emitting plasma is changing making the plasma more luminous. Therefore, the changing $\dot{m}_{\rm acc}$ does not necessarily describes real $\dot{m}_{\rm acc}$ evolution.

If we still assume that the small value of $\dot{m}_{\rm acc}$ is real because only a small fraction of $\dot{m}_{\rm acc}$ is evaporated into the corona, the majority of the accreted mass should flow through the underlying geometrically thin disc, and the corresponding boundary layer should be optically thick and it should be seen as a soft component in the X-ray spectra. We did not detect it. As mentioned in Section~\ref{introduction}, only very few DNe show this optically thick boundary layer, and similar search for this soft component in nova-like systems was not successful. Apparently, the existence of an optically thick boundary layer in the high state of CVs is questionable even if the theory predicts it.

\section{Summary and Conclusions}

We observed the DNe V1504\,Cyg with \xmm\ during the decline from an outburst covering a specific transition phase during which the X-ray flux, amplitude of the variability and the hardness ratio have risen in a very short time interval. The UV light curve did not follow this trend, it only decreased in flux and amplitude of the variability. This different behaviour of X-ray and UV light curve we explain by truncated inner disc and a missing reprocessing region.

Spectral fitting assuming a cooling flow model revealed higher temperatures and mass accretion rate during the second part of the observation when the X-ray flux was higher. The first part of the data with lower X-ray flux is better described by an additional power law component. This component should represent the scattering in an outflow, which was already observed in nova-like systems in the high state (\citealt{balman2014}), or thermal Comptonization of the soft disk photons. The rising mass transfer rate during the decline from an outburst derived with spectral modeling reflects the changing radiation efficiency of the evaporating inner disc and not the real increase of the mass flow. Such radiation inefficiency is typical for ADAFs.

Timing analysis of the observation showed a flat PDS during the low flux part of the observation, while the second part with higher flux showed a red noise shape with a potential break frequency at log($f$/Hz) = -3.02 with confidence approximately 2-$\sigma$. This detection agrees with the break frequency detected in optical by \citet{dobrotka2015} and \citet{dobrotka2016}. The authors proposed that this frequency is generated by the innermost parts of the accretion disc. The X-ray nature of the variability found in this work imply a connection with the central hot optically thin plasma. Moreover, similar frequency was observed in several other cataclysmic variables during the high state, while it was missing during the low state. Based on the reprocessing geometry, a sandwich model where a geometrically thick ADAF-like corona surrounds the geometrically thin disc is a possible accretion configuration.

\section*{Acknowledgment}

AD was supported by the Slovak grant VEGA 1/0408/20, and by the European Regional Development Fund, project No. ITMS2014+: 313011W085. We acknowledge with thanks the variable star observations from the AAVSO International Database contributed by observers worldwide and used in this research. This work makes use of ASAS-SN (\citealt{shappee2014}, \citealt{kochanek2017}) and \kepler\ data (\citealt{borucki2010}). We thank the anonymous referee for very helpful comments.

\bibliographystyle{aa}
\bibliography{mybib}

\label{lastpage}

\end{document}